\documentclass[journal]{IEEEtranTIE}

\usepackage{graphicx}
\usepackage{cite}
\usepackage{amsmath}
\usepackage{url}
\usepackage{colortbl}
\usepackage{soul}
\usepackage{multirow}
\usepackage{pifont}
\usepackage{color}
\usepackage{alltt}
\usepackage[hidelinks]{hyperref}
\usepackage{enumerate}
\usepackage{siunitx}
\usepackage{epstopdf}
\usepackage{pbox}
\usepackage{comment}
\usepackage{mathtools}
\usepackage{adjustbox}

\begin{document}

\title{A Direct Mapped Method for Accurate Modeling and Real-Time Simulation of High Switching Frequency Resonant Converters}

\author{
	\vskip 1em
	{H. Chalangar, \emph{Student Member}, T. Ould-Bachir, \emph{Member, IEEE}, \\K. Sheshyekani, \emph{Senior Member, IEEE}, and J. Mahseredjian, \emph{Fellow Member, IEEE}}

	\thanks{
		
		{Manuscript received January 30, 2020; revised April 13, 2020; accepted May 14, 2020.
		
		H. Chalangar, K. Sheshyekani and J. Mahseredjian are with the Department of Electrical Engineering, Polytechnique Montréal, H3T 1J4, Qc, Canada (e-mail: hossein.chalangar@polymtl.ca; keyhan.sheshyekani@polymtl.ca; jean.mahseredjian@polymtl.ca).
		
		T. Ould Bachir is with the Department of Computer and Software Engineering, Polytechnique Montréal, H3T 1J4, Qc, Canada  (e-mail: tarek.ould-bachir@polymtl.ca).
		}
	}
}

\maketitle

\begin{abstract}

This paper presents a Direct Mapped Method (DMM) for real-time simulation of high switching frequency resonant converters. The DMM links state variables to diode statuses and provides an exact and noniterative solution to network equations. The proposed method is implemented on FPGA to simulate an LLC converter with switching frequencies ranging up to 500~kHz. The best reported implementations of DMM achieve a 25~ns simulation time-step for a wide range of clock frequencies, ranging from 40~MHz to 320~MHz. 

\end{abstract}


\markboth{IEEE TRANSACTIONS ON INDUSTRIAL ELECTRONICS}%
{H. Chalangar \MakeLowercase{\textit{et al.}}: A Direct Mapped Method}

\definecolor{limegreen}{rgb}{0.2, 0.8, 0.2}
\definecolor{forestgreen}{rgb}{0.13, 0.55, 0.13}
\definecolor{greenhtml}{rgb}{0.0, 0.5, 0.0}

\section{Introduction}

\IEEEPARstart{A}{Advances} in semiconductor technology allow the increase of the efficiency and power density of modern power electronic converters (PECs) by operating them at higher switching frequencies ($>$50 kHz). Resonant converters are particularly of interest and renowned for their higher power density with typically small L/C components~\cite{Pilawa2009,PontLLC2019}. These converters are typically operated using soft switching techniques to preclude switching losses, reduce filtering requirements, and to mitigate electromagnetic interference (EMI)~\cite{Bellar1998, Fang2013}. Among various resonant topologies~\cite{Ericksonbook}, the LLC converter has drawn attention for its unique characteristics such as low voltage stress on the secondary rectifier and high efficiency at high input voltages~\cite{Yang2002, Xie2007}.

These technology advancements are however very challenging for the realm of FPGA-based Hardware-In-the-Loop (HIL) real-time simulation (RTS). HIL simulation is a prototyping technique used to assess the performance of  control/protection systems in the design or installation stages. FPGAs are programmable devices that have played a pivotal role during the last decade in the RTS of PECs for HIL applications~\cite{Matar2010, Tarek2013, Dagbagi2016}. The main challenges faced by the FPGA-based HIL of high switching frequency (HSF) converters come from the limited memory resources and the heavy processing requirements given the small calculation time-step. 

Two main modeling approaches are typically employed in FPGA-based RTS to model switching devices: i) The Associate Discrete Circuit (ADC); and ii) The Resistive Switch Model (RSM). The ADC switch model was introduced in the 90s~\cite{Pejovic1994} and has been since extensively utilized for RTS applications, provided that the simulation time-step ($\Delta t$) is sufficiently small\cite{Matar2010, Tarek2013, Dagbagi2016}. However, ADC is prone to fictitious oscillations that can alter the behavior of a PEC model in particular when dealing with HSF. Various techniques exist to alleviate these drawbacks~\cite{RAZZAGHI2014,GenerelizedADC2019, Mu2014, DufourPatent}, but are unpractical for resonant converters because the adopted switch model affects the resonant tank behaviour. The RSM on the other hand uses a two-valued resistance for the switch ($R_\textit{on}\simeq 0$ and $R_\textit{off}\gg 0$)~\cite{Blanchette2012,Tarek2015,Hadizadeh2019, FeiGao2019, Iravni2019}. RSM produces more accurate results compared to ADC when modeling PECs, but has the drawbacks of a variable admittance matrix and heavier computations in real-time.

To circumvent the heavy computational burden of RSM, the inverse of the nodal matrix for all possible switch combinations can be precomputed. This approach, however, considerably limits the number of switches due to FPGA memory requirement limitations. The Sherman-Morrison-Woodbury formula has been utilized to compute the updated inverse matrix on the fly~\cite{Hadizadeh2019}, but the method is unpractical for HSF PECs, and neglects the iterations needed to determine the state of natural commutation devices such as diodes. Methods for handling the RTS of diodes can be found in~\cite{Blanchette2012, Tarek2015, Iravni2019, FeiGao2019}. In~\cite{Blanchette2012}, unrealistic parasitic elements are augmented in parallel with switches that may alter the behavior of the converter, more so when resonant converters are considered. The method proposed in~\cite{Tarek2015} results in large time-steps and as such is inadequate for the simulation of HSF PECs. A predictor-corrector algorithm has been used in~\cite{FeiGao2019} to decouple the switches from the circuit elements and to simulate them simultaneously. However, due to the stability concerns, very short time-steps should be selected which results in increased hardware resources when dealing with large circuits. More recently, the RSM has been used to simulate low-frequency PECs~\cite{Iravni2019}. This method, however, uses an iterative solver to deal with switches, which makes it unpractical for the simulation of HSF PECs.

The RTS of HSF converters is more challenging since it requires very small time-steps which can be restrictive for the iterative solver. To obviate the iterative procedure while adopting the RSM, some specific simplifications are often made for the determination of switch states. In \cite{SiemaszkoLLC2014}, a commercial RTS method has been used to simulate an LLC, with the switching frequency of 20~kHz which is relatively low. The simulation of an LLC converter with a switching frequency of 60~kHz has been presented in \cite{JiLLC2016}, where the LLC model is built by breaking down the converter into interconnected modules. This approach relies on intra-module delays which can cause erroneous results. In \cite{FeiGaoLLC2019}, an LLC converter with a resonance frequency of 160~kHz has been simulated using FPGA. The simulations have been done by making use of Forward Euler (FE) method where a time-step of 15~ns has been achieved. The FE is not however a generic approach and might not work for certain LLC parameters. This will be discussed for the first time in this paper. 

This paper proposes a Direct Mapped Method (DMM) for accurate RSM-based simulation of HSF resonant converters. The method offers higher computational performance while providing the same level of accuracy compared to iterative solutions. The DMM proceeds by constructing a mapping function that relates the state variables of the circuit to switch states. The mapping function is then used to directly determine switch states during the simulation. In this paper, the DMM is used to implement an FPGA-based LLC simulator with parameters taken from the literature. It is demonstrated that the method is successful in the accurate simulation of LLC converters with switching frequencies ranging up to 500~kHz.

The remainder of this paper is organized as follows. The theoretical background of resonant and LLC converters are briefly presented in Section~\ref{sec:LLC}. In Section~\ref{sec:DMM}, the DMM is elaborately presented while its application for an LLC converter is discussed. Section~\ref{sec:FPGA} presents the FPGA implementation of the DMM and discusses experimental results.

\section{DC-DC Resonant Converter }
\label{sec:LLC}

\begin{figure}[!t]
    \centering
    \includegraphics [width=3.4 in] {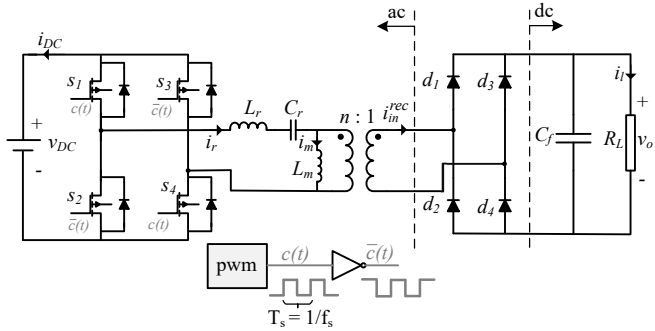}
    \caption{LLC converter used in this paper.}
    \label{fig:LLC circuit}
\end{figure}

A typical DC-DC resonant converter is comprised of three stages: an inverter, a 2- or 3-element resonant tank, and a rectifier. The inverter consists of a MOSFET half-bridge or full-bridge, operated with a fixed 50\% duty cycle. The power flow of the converter is controlled by modulating the frequency of the square wave with respect to the tank's resonant frequency. The rectifier stage converts its AC input to a DC output voltage which is filtered by a capacitor. A transformer is often used in the rectification stage for scaling and isolation purposes.

The resonant tank modulates the voltage using L-C elements such as series LC, parallel LC, LLC, etc. These resonant tanks are selected depending on the converter application~\cite{ClassificationofresonanrHuang2011}. In this work, we will focus on the LLC converter topology which is comprised of a full-bridge inverter and a full-bridge rectifier, see Fig.~\ref{fig:LLC circuit}. The LLC converter is used in many applications such as power electronic-based distributed generation, electric vehicles, computer and communication systems \cite{Sun2015, Zubieta2015,Haoyu2014, 300kHzLLC}. The proposed DMM is, however, applicable to other resonant topologies as well.

\begin{figure}[!t]
    \centering
    \includegraphics [width=3.3 in] {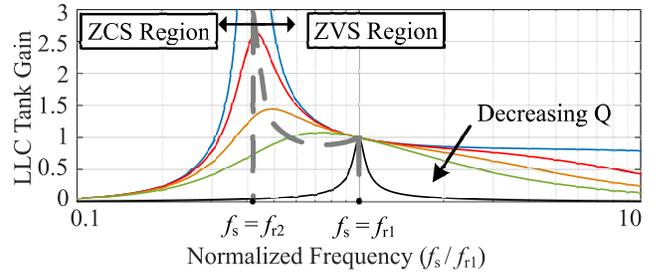}
    \caption{Voltage gain of LLC tank versus $F = f_s/f_r$ for different quality factors.}
    \label{fig:LLC gain}
\end{figure}

\subsection{Overview of the LLC Resonant Converter}

The LLC converter is a multi-resonant converter with resonant frequencies $f_{r}=f_{r_1}=1/(2\pi \sqrt{L_r C_r})$ and $f_{r_2}=1/(2\pi \sqrt{(L_r+L_m)C_r})$. With reference to Fig.~\ref{fig:LLC circuit}, $L_m$ is embodied in the magnetizing inductance of the transformer which helps to reduce the size and the cost of the the converter\cite{Zubieta2015}. LLC resonant tank gain, $G(F)$, is defined as the magnitude of its input–output voltage transfer function. Typical voltage gains of the LLC tank for different normalized switching frequencies $F=f_s/f_r$, ($f_s$ being the switching frequency) are obtained from~\cite{infenion2}:

\begin{equation}
G(F) = \dfrac{(m-1)F^2}{\sqrt{(m F^2-1)^2 + (m-1)^2F^2(F^2-1)^2Q^2}},
\label{LLC gain}
\end{equation}
where $Q$ is the quality factor, and $m$ is the inductance ratio:

\begin{equation}
\left \{
\begin{array}{l}
Q = (\sqrt{L_r/C_r})/R_\mathit{ac} \\
m = (L_r + L_m)/L_r\\ 
R_\mathit{ac} = (8n^2/\pi ^2)R_L \\ 
\end{array}
\right.~~.
\label{LLC parameter defination}
\end{equation}

$R_\mathit{ac}$ denotes the reflected load resistance $R_L$, seen from the resonant tank side, and $n$ is the turn ratio of the isolation transformer.

\subsection{LLC Operation}
\label{Section : LLC simulation} 

Fig.~\ref{fig:LLC gain} depicts the LLC voltage gain for different values of quality factor $Q$, which are used for the analysis and design of the resonant converter. $f_{r_2}$ delimits the capacitive and the inductive regions of the resonant tank, which are associated with the Zero Current Switching (ZCS) and Zero Voltage Switching (ZVS) regions, respectively. The MOSFET-based LLC converters are preferably operated in the ZVS region, i.e. $f_s>f_{r_2}$, since it significantly mitigates the switching losses\cite{Ericksonbook}. When LLC is working with $f_{r_2}<f_s <f_{r_1}$, at the end of one half resonant cycle, during a certain time interval no power is delivered to the load and the whole resonant current passes through the magnetizing inductance. During this time interval, the rectifier is said to be in blocking mode ($i_\textit{in}^\textit{rec} = n|i_r - i_m| \simeq 0$). The region where $f_s > f_{r_1}$ refers to an operational mode of the LLC where a resonant half cycle is not completed and interrupted by the switching of other MOSFET.

\subsection{LLC Simulation}
\label{Subsection : LLC simulation}

In this paper, the LLC converter of Fig.~\ref{fig:LLC circuit} is considered for which two sets of parameters are chosen from the literature as listed in Table~\ref{tab:LLCDATA}. In this table, LLC with Parameter Set \#1 \cite{FeiGaoLLC2019} and Parameter Set \#2 \cite{Fei2018} has the resonant frequency of 160 kHz, 500 kHz, respectively. We will show in this section why an iterative solution is needed to accurately simulate the LLC converter. 

As proposed in \cite{FeiGao2019} and \cite{FeiGaoLLC2019}, an explicit integration method such as FE can be used to avoid the iterative process. This however necessitates adopting a sufficiently small simulation time-step. For the sake of clarity, let us recall that FE offers a recursive solution to $\dot{x}(t) = f(x(t))$ in the following form: 

\begin{equation}
x(t) = x(t-\Delta t) + \Delta t f(x(t-\Delta t)). 
\label{FE formula}
\end{equation}

By applying Eq.~(\ref{FE formula}) to the resonant inductors of the resonant tank of the LLC, an explicit expression for the rectifier's input current is obtained (see Fig.~\ref{fig:LLC circuit}):

\begin{equation}
\begin{array}{ll}
    i_\textit{in}^\textit{rec}(t) &= n\{i_r(t)-i_m(t)\} \\ [0.3em]
    &=n\{i_r(t-\Delta t)+\frac{\Delta t}{L_r} v_{L_r}(t-\Delta t) \\ [0.3em]
    &-i_m(t-\Delta t)-\frac{\Delta t}{L_m} v_{L_m}(t-\Delta t) \}.
\end{array}
\label{FELLC}
\end{equation}

\begin{figure}[!t]
    \centering
    \includegraphics [width=3.3 in] {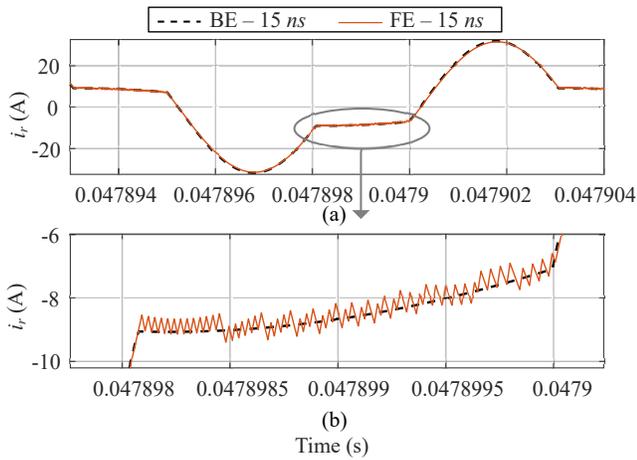}
    \caption{LLC with Parameter Set~\#1: (a)-Resonant current ($i_r$) of the LLC converter calculated by FE and BE methods with identical time-step of 15~ns, (b)-Close-up view of $i_r$ during blocking mode.}
    \label{fig:ircase1}
\end{figure}

It is proposed in~\cite{FeiGaoLLC2019} to use Eq.~(\ref{FELLC}) to determine the mode of the full-bridge rectifier; positive mode when $i_\textit{in}^\textit{rec}(t) \ge 0$, $(d_1, d_2, d_3, d_4)=(1, 0, 0, 1)$, and negative mode otherwise --- $(d_1, d_2, d_3, d_4)=(0, 1, 1, 0)$. However, such an approach may lead to either inaccurate results or instability, even for very small time-steps. This behavior is partly due to the fact that this method completely ignores the blocking mode of the converter.

\begin{figure}[!t]
    \centering
    \includegraphics [width=3.3 in] {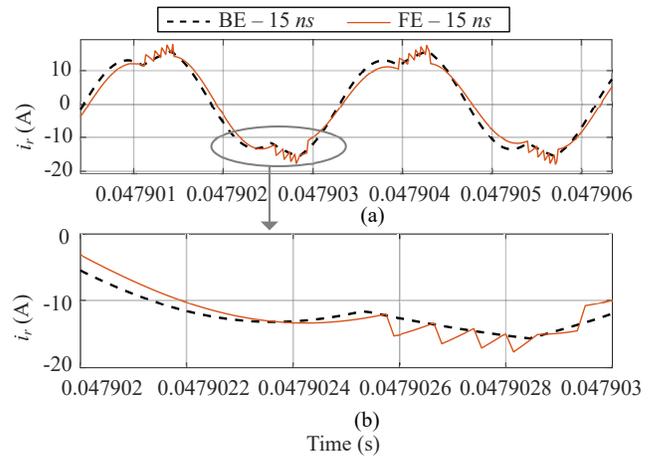}
    \caption{LLC with Parameter Set~\#2: (a)-Resonant current ($i_r$) of the LLC converter calculated by FE and BE methods with identical time-step of 15~ns, (b)-Close-up view of $i_r$ during blocking mode.}
    \label{fig:ircase3}
\end{figure}

\begin{table}[!t]
\centering
\caption{Circuit Parameters for LLC Circuits}
\label{tab:LLCDATA}
\begin{tabular}{|c|c|c|}
\hline
Parameter              & Parameter Set \#1\cite{FeiGaoLLC2019} &  Parameter Set \#2 \cite{Fei2018} \\ \hline
$f_{r_1}$ (kHz)        &  160                            &  500                       \\
$f_{r_2}$ (kHz)        &  60                             &  210                       \\
$n$:1                  &  1.5:1                          &  33:1                      \\
$V_{\textit{in}}$ (V)  &  600                            &  400                       \\
$V_{\textit{out}}$ (V) &  400                            &  12                        \\
$L_r$ ($\mu$H)         &  25                             &  4.5                       \\
$L_m$ ($\mu$H)         &  150                            &  21.6                      \\
$C_r$ (nF)             &  40                             &  22                        \\
$C_o$ ($\mu$F)         &  1000                           & 3000                       \\
P(kW)                  &  5.3                            &  1                         \\
$R_L$($\Omega$)        &  30                             &  0.144                     \\
\hline
\end{tabular}
\end{table}

Fig.~\ref{fig:ircase1}.a gives the resonant current, $i_r$, for LLC with Parameter Set \#1 (see Table ~\ref{tab:LLCDATA}), calculated by either making use of FE or backward Euler (BE) methods, and assuming a time-step of 15~ns. It is noted that BE method is applied through an iterative solution. The zoomed view of resonant current during the blocking mode is shown in Fig.~\ref{fig:ircase1}.b. It is seen from this figure that in contrast with BE method, the current $i_r$ calculated by the method proposed in~\cite{FeiGaoLLC2019} is oscillatory during the blocking mode period. The blocking mode is in fact disregarded during the simulation i.e., only positive and negative modes are considered, and the model alternates between the two modes during the blocking period. 

The effects of ignoring the blocking mode for LLC with Parameter Set~\#2 (see Table ~\ref{tab:LLCDATA}) are observed in Fig.~\ref{fig:ircase3}. Similar to the previous case, $i_r$ is calculated by either BE or FE with the same 15~ns time-step. It is seen from Fig.~\ref{fig:ircase3} that $i_r$ calculated by FE is oscillatory during the blocking mode. It is also seen from Fig.~\ref{fig:ircase3} that, as opposed to LLC with Parameter Set~\#1, the results of FE and BE are quite different for LLC with Parameter Set~\#2. This indicates that the accuracy of the method proposed in~\cite{FeiGaoLLC2019} can be affected by circuit parameters. 

During our simulations, it was also observed that the resonant current calculated by FE method becomes unstable for $\Delta t \ge30$~ns. It is also worth noting that another possible mode of the rectifier stage of the LLC circuit of Fig.~\ref{fig:LLC circuit} is the short circuit mode, i.e. $(d_1, d_2, d_3, d_4)=(1, 1, 1, 1)$ which is also disregarded by explicit methods such as\cite{FeiGaoLLC2019, FeiGao2019}.

\section{Direct Mapped Method}
\label{sec:DMM}

\subsection{DMM Analysis of the LLC}

\begin{figure}[!t]
    \centering
    \includegraphics [width=3.4 in] {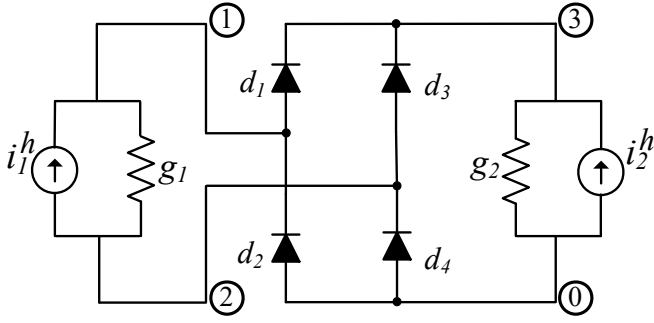}
    \caption{Single-phase rectifier circuit.}
    \label{fig:Single-Phase rectifier}
\end{figure}

Without losing any generality, we consider the circuit of Fig. \ref{fig:Single-Phase rectifier}, which represents the rectifying stage of the LLC. Both AC and DC side circuits are replaced by a Norton equivalent circuit comprised of a conductance in parallel with a time-dependent current source. The Norton equivalents result from the BE discretizition. 

The following systems of equations associate network equations of the circuit of Fig.~\ref{fig:Single-Phase rectifier} using a classical nodal analysis:

\begin{equation}
\mathbf{Y}^{\sigma_\textit{rec}}\mathbf{v}_n=\mathbf{i}
~,
\label{Eq:system of equation 1}
\end{equation}
\begin{equation}
\resizebox{.91\hsize}{!}{$
\mathbf{Y}^{\sigma_\textit{rec}}=
\begin{bmatrix*}[l]
g_1+g_{d_1}+g_{d_2}  & -g_1                 & -g_{d_1}\\
-g_1                 & g_1+g_{d_3}+g_{d_4}  & -g_{d_3}\\
-g_{d_1}             & -g_{d_3}             & g_2+g_{d_1}+g_{d_3}
\end{bmatrix*}$,}
\label{AdmittanceMatrixByelements}
\end{equation}
\begin{equation}
\mathbf{i}=
\begin{bmatrix}
i^h_{1}, & -i^h_{1}, & i^h_{2} 
\end{bmatrix},
\label{currentInjectionU}
^T
\end{equation}

\noindent where $\mathbf{Y}^{\sigma_\textit{rec}}$ is the admittance matrix, $\sigma_\textit{rec}$ $\in$ $\{0,1,...,15\}$ refers to the 16 diodes status combinations, $\mathbf{v}_n=\lbrack v_{1}, v_{2}, v_{3}\rbrack ^T$ is the unknown nodes voltages and $\mathbf{i}$ is the current injection vector. $g_{d_i}=1/R_{d_i}$, $i \in \{1, 2, 3, 4\}$ are the conductances associated with the diodes ($g_{d_i} = g_\textit{on}$ or $g_\textit{off}$, with respect to the diode status combination $\sigma_\textit{rec}$).

Eq.~(\ref{Eq:system of equation 1}) can be rewritten as:

\begin{equation}
\mathbf{Y}^{\sigma_\textit{rec}}\mathbf{v}_{n}=\mathbf{i}=\mathbf{B}\mathbf{i^h}
~,
\label{Eq:system of equation 2}
\end{equation}

\begin{equation}
\mathbf{B}
=
\begin{bmatrix}
+1 & 0 \\
-1 & 0 \\
 0 & +1
\end{bmatrix}
\label{Bieq}.
\end{equation}

\begin{equation}
\mathbf{i}^h = \lbrack i^h_{1}, i^h_{2}\rbrack ^T,
\label{ih}
\end{equation}

In addition, diodes voltages can be defined using a connectivity matrix, $\mathbf{T}$, as:

\begin{equation}
\mathbf{v_d}=\mathbf{T}\mathbf{v}_n
~,
\label{Connectitivity Diode Voltage}
\end{equation}
\begin{equation}
\mathbf{T}=
\begin{bmatrix}
+1 & 0 & -1\\
-1 & 0 & 0 \\
 0 & +1 & -1 \\
 0 & -1 & 0
\end{bmatrix}
\label{define-T}
\end{equation}

\noindent where $\mathbf{v_d}$ is the vector of diode voltages: $\mathbf{v_d} = \lbrack v_{d_1}, v_{d_2}, v_{d_3}, v_{d_4}\rbrack ^T$. From (\ref{Eq:system of equation 2}) and (\ref{Connectitivity Diode Voltage}), $\mathbf{v_d}$ can be obtained as follows:

\begin{equation}
\mathbf{v_d} = \mathbf{T}(\mathbf{Y}^{\sigma_\textit{rec}})^{-1}\mathbf{B}\mathbf{i^h}.
\label{Diode volatge in matrx form}
\end{equation} 

The RSM requires the correct state of each diode (ON/OFF) at each time-point. This is done by checking whether the sign of the diode's voltage or its current ($v_{d_i}= R_{d_i}i_{d_i},~R_{d_i}>0$) is positive or negative. However, from (\ref{Diode volatge in matrx form}), one can find that diode $d_i$ conducts whenever $\vartheta_{d_i}>0$, $i\in\{1, 2, 3, 4\}$:  

\begin{equation}
\begin{array}{ll}
\vartheta_{d_1} = &+(g_{d_2}g_{d_3} + g_{d_3}g_{d_4} + g_{d_3}g_{2} + g_{d_4}g_{2})i^h_{1} \\ [.2em] 
& - (g_{d_2}g_{d_3} + g_{d_2}g_{d_4} + g_{d_2}g_{1} + g_{d_4}g_{1})i^h_{2}\\ [.2em] 
\end{array}~,
\label{d1 voltage equation}
\end{equation} 

\begin{equation}
\begin{array}{ll}
\vartheta_{d_2} = &-(g_{d_1}g_{d_4} + g_{d_3}g_{d_4} + g_{d_3}g_{2} + g_{d_4}g_{2})i^h_{1} \\ [.2em] 
& - (g_{d_1}g_{d_3} + g_{d_1}g_{d_4} + g_{d_1}g_{1} + g_{d_3}g_{1})i^h_{2}\\[.2em]
\end{array}~,
\label{d2 voltage equation}
\end{equation} 

\begin{equation}
\begin{array}{ll}
\vartheta_{d_3} = &-(g_{d_1}g_{d_2} + g_{d_1}g_{d_4} + g_{d_1}g_{2} - g_{d_2}g_{2})i^h_{1} \\ [.2em] 
& - (g_{d_1}g_{d_4} + g_{d_2}g_{d_4} + g_{d_2}g_{1} + g_{d_4}g_{1})i^h_{2}\\[.2em]
\end{array}~,
\label{d3 voltage equation}
\end{equation} 

\begin{equation}
\begin{array}{ll}
\vartheta_{d_4} = &+(g_{d_1}g_{d_2} + g_{d_2}g_{d_3} + g_{d_1}g_{2} + g_{d_2}g_{2})i^h_{1} \\ [.2em] 
& - (g_{d_1}g_{d_3} + g_{d_2}g_{d_3} + g_{d_1}g_{1} + g_{d_3}g_{1})i^h_{2}\\ [.2em]
\end{array}.
\label{d4 voltage equation}
\end{equation} 

\begin{figure}[!t]
\centering
\includegraphics[width=3.3 in]{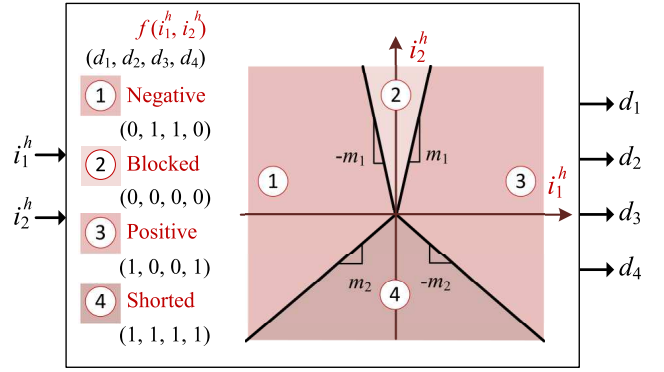}
\caption{Mapping history currents to diode statuses.}
\label{fig:Mapping function for rectifier}
\vspace{-1em}
\end{figure}

Hence, for each combination $\sigma_\textit{rec}$, a system of inequalities combining (\ref{d1 voltage equation})-(\ref{d4 voltage equation}) is obtained, and the Fourier-Motzkin elimination~\cite{MatouekFourierMotzkin} is  used to evaluate its feasibility. 

The feasibility check reveals that only 4 feasible diode combinations exist, those for which $d_1 = d_4$ and $d_2 = d_3$. A mapping function $f(i_1^h, i_2^h)$ linking the state variables to diode states, $(i_1^h, i_2^h)\longmapsto \sigma_\textit{rec} \equiv
(d_1, d_2, d_3, d_4)$, can be obtained.  Fig.~\ref{fig:Mapping function for rectifier} demonstrates this mapping function in the $i_1^hi_2^h$-plane with four half lines dividing the plane into four feasible regions: (0, 0, 0, 0) = Blocked, (1, 0, 0, 1) = Positive, (0, 1, 1, 0) = Negative and (1, 1, 1, 1) = Shorted. The half lines start at the origin and have slopes of $\pm m_1$ and $\pm m_2$, as illustrated in Fig.~\ref{fig:Mapping function for rectifier}. The slopes $m_1$ and $m_2$ are given by:

\begin{equation}
m_1 = \frac{g_{\textit{off}}g_{\textit{off}} + g_{\textit{off}}g_{\textit{off}} + g_{\textit{off}}g_{2} + g_{\textit{off}}g_{2}}{g_{\textit{off}}g_{\textit{off}} + g_{\textit{off}}g_{\textit{off}} + g_{\textit{off}}g_{1} + g_{\textit{off}}g_{1}},
\label{m1 equation}
\end{equation} 

\begin{equation}
m_2 = \frac{g_{\textit{on}}g_{\textit{on}} + g_{\textit{on}}g_{\textit{on}} + g_{\textit{on}}g_{2} + g_{\textit{on}}g_{2}}{g_{\textit{on}}g_{\textit{on}} + g_{\textit{on}}g_{\textit{on}} + g_{\textit{on}}g_{1} + g_{\textit{on}}g_{1}}.
\label{m2 equation}
\end{equation} 

Eqs.~(\ref{m1 equation}) and (\ref{m2 equation}) reduce to:

\begin{equation}
m_1 = \dfrac{g_\textit{off} + g_{2} }{g_\textit{off} + g_{1}},
\label{m1 simple}
\end{equation} 

\begin{equation}
m_2 = \dfrac{g_\textit{on} + g_{2} }{g_\textit{on} + g_{1}}.
\label{m2 simple}
\end{equation} 

Assuming $g_\textit{off} \simeq 0$, slope $m_1$ is approximated by:

\begin{equation}
m_1  \simeq \dfrac{g_2}{g_1}.
\label{Slopes of DMM1}
\end{equation}

From (\ref{Slopes of DMM1}), one can observe that the slope $m_1$ (which delimits the Blocked from the Positive and Negative states) is approximately independent of diode conductances and is a function of both AC and DC side parameters. It also appears from Eq.~(\ref{Slopes of DMM1}) that explicit solvers such as those used in \cite{FeiGao2019,FeiGaoLLC2019} --- which substitute the Norton equivalent of the AC side by a pure current source ($g_1 = 0$) --- discard the Blocked mode of the rectifier because they force $m_1$ to infinity. Such an approximation is not problematic if the slope $m_1$ is very large such as for Parameter Set~\#1, but can considerably degrade the simulation results for other parameters, such as those of Parameter Set~\#2.

\subsection{Simulation Algorithm}
\label{sec:algo}

The aim of this section is to develop a simulation algorithm for the LLC converter of Fig.~\ref{fig:LLC circuit}. The inverter is operated in controlled mode, and DMM is used to determine the diode states. Network equations for the LLC converter of Fig.~\ref{fig:LLC circuit} are formulated using the Modified Augmented Nodal Analysis (MANA)~\cite{JeanEMTPpaper} after discretizing all elements using BE rule:

\begin{equation}
  \mathbf{A}^{\sigma}\mathbf{x}(t)= \mathbf{B}
   \begin{bmatrix} 
    u(t),    \mathbf{i}^\textit{hist}_{\textrm{ac}}(t), 
   \mathbf{i}^\textit{hist}_{\textrm{dc}}(t)
   \end{bmatrix}^T.
   \label{MANA}
\end{equation}

\noindent where $\mathbf{A}^{\sigma}$ is the MANA matrix for switch combination $\sigma$, $\mathbf{B}$ is the incidence matrix, $\mathbf{x}(t)$ is the vector of unknown variables. $u(t) = v_\textrm{dc}(t)$ is the input DC voltage, $\mathbf{i}^\textit{hist}_{\textrm{ac}}(t)$, and $\mathbf{i}^\textit{hist}_{\textrm{dc}}(t)$ are ac and dc side history vectors, respectively. $\mathbf{x}(t)$ is comprised of node voltages ($\mathbf{v}_n(t)$), the current entering the DC voltage source ($i_\textrm{DC}(t)$) and the current entering the secondary port of the transformer ($i_\textrm{DB}(t) = -i_\textit{in}^\textit{rec}(t)$): $\mathbf{x}(t)=[\mathbf{v}_n(t), i_\textrm{DC}(t), i_\textrm{DB}(t)]$. The history vectors are comprised of the history currents that result from the backward discretization of the L/C components in the circuits: $\mathbf{i}^\textit{hist}_{\textrm{ac}}(t)=[i^\textit{hist}_{C_r}(t), i^\textit{hist}_{L_r}(t), i^\textit{hist}_{L_m}(t)]$, and $\mathbf{i}^\textit{hist}_{\textrm{dc}}(t)=[i^\textit{hist}_{C_f}(t)]$. More on the use of MANA for the simulation of power electronic circuits can be found in~\cite{Tarek2015, montano2018}. 

To reduce the computational burden of solving Eq.~(\ref{MANA}) at each time-point of the simulation, the following formulation is used:

\begin{equation}
   \begin{bmatrix} 
   \mathbf{i}^\textit{hist}_{\textrm{ac}}(t+\Delta t) \\[0.3em]
   \mathbf{i}^\textit{hist}_{\textrm{dc}}(t+\Delta t)\\[0.3em]
   \mathbf{y}(t)
   \end{bmatrix} 
   = 
   \begin{bmatrix} 
   \mathbf{H}^{\sigma(t)}_\textrm{ac,u}  & \mathbf{H}^{\sigma(t)}_\textrm{ac,ac} & \mathbf{H}^{\sigma(t)}_\textrm{ac,dc}   \\[0.2em]
   \mathbf{H}^{\sigma(t)}_\textrm{dc,u}  & 
   \mathbf{H}^{\sigma(t)}_\textrm{dc,ac} & 
   \mathbf{H}^{\sigma(t)}_\textrm{dc,dc}   \\[0.2em]
   \mathbf{H}^{\sigma(t)}_\textrm{y,u}   &
   \mathbf{H}^{\sigma(t)}_\textrm{y,ac}  &
   \mathbf{H}^{\sigma(t)}_\textrm{y,dc}
   \end{bmatrix} 
      \begin{bmatrix} 
   u(t) \\[0.3em]
   \mathbf{i}^\textit{hist}_{\textrm{ac}}(t) \\[0.3em]
   \mathbf{i}^\textit{hist}_{\textrm{dc}}(t)
   \end{bmatrix}~~, 
   \label{Eq:W}
\end{equation}

\noindent where $\mathbf{H}^\sigma_\textrm{-,-}$ are precomputed matrices obtained from simple algebraic manipulations of $\mathbf{B}$ and inverse of $\mathbf{A}^\sigma$. $\mathbf{y}(t)$ is a vector comprised of desired output variables, $\mathbf{y}(t)=[v_o(t), i_r(t), i_m(t)]$. Similar rewritings are used to determine the history currents associated with the Norton equivalents of the ac and dc sides: 

\begin{equation}
\left\{
\begin{array}{ll}
i^h_1(t)
   = &  
   \mathbf{H}^{\sigma_\textit{inv}(c(t))}_{1\textrm{,u}}u(t) + \mathbf{H}^{\sigma_\textit{inv}(c(t))}_{1\textrm{,ac}}\mathbf{i}^\textit{hist}_{\textrm{ac}}(t) \\[0.3em]
i^h_2(t)
   = &\mathbf{H}_2   \mathbf{i}^\textit{hist}_{\textrm{dc}}(t)  
\end{array}
\right.~~,
   \label{Eq:W1 and W2}
\end{equation}
where $\sigma_\textit{inv}$ is the switch combination associated with the inverter's state, which is defined as a function of the gating signal $c(t)$ driving $S_1$ and $S_4$ (see Fig.~\ref{fig:LLC circuit}), given the controlled operation of the inverter:
\begin{equation}
\sigma_\textit{inv}(c(t)) = \left \{
\begin{array}{ll}
6, & \text{if~} c(t) = 0 \\
9, & \text{if~} c(t) = 1
\end{array}
\right.~.
\label{Eq:sigmainv}
\end{equation}

The DMM function is used to determine $\sigma_\textit{rec}$ and $\sigma$:

\begin{equation}
    \sigma_\textit{rec}(t) = f(i^h_1(t), i^h_2(t))\text{,~} \sigma_\textit{rec}(t) \in \{0, 6, 9, 15\},
    \label{Eq:sigmarec}
\end{equation}

\begin{equation}
\sigma(t) = 16\cdot\sigma_\textit{inv}(c(t)) + \sigma_\textit{rec}(t)  .
\label{Eq:sigma}
\end{equation}

DMM consists mainly in solving the following (see Fig.~\ref{fig:Mapping function for rectifier}):

\begin{equation}
\begin{bmatrix}
p_1(t) \\
p_2(t) \\
p_3(t) \\
p_4(t)
\end{bmatrix} = \text{sgn}\left(
\begin{bmatrix}
-m_1 & 1\\
+m_1 & 1\\
-m_2 & 1\\
+m_2 & 1
\end{bmatrix}
\begin{bmatrix}
i^h_1(t) \\
i^h_2(t)
\end{bmatrix}\right),
\label{eq:solve f}
\end{equation}
where $\text{sgn}(\cdot)$ is the sign function. The DMM function reads henceforth as follows:

\begin{equation}
\left\{
\begin{array}{l}
(p_1(t),~p_2(t)) = (+1, +1) \Rightarrow \sigma_\textit{rec}(t) = 0 \\
(p_2(t),~p_3(t)) = (-1, +1) \Rightarrow \sigma_\textit{rec}(t) = 6 \\
(p_1(t),~p_4(t)) = (-1, +1) \Rightarrow \sigma_\textit{rec}(t) = 9 \\
(p_3(t),~p_4(t)) = (-1, -1) \Rightarrow \sigma_\textit{rec}(t) = 15
\end{array} 
\right..
\label{Eq:pfct}
\end{equation}

From the mathematical formulation given above, the following algorithm is used at each time-point to simulate the LLC:

\begin{enumerate}
   \item Determine $\sigma_\textit{rec}(t)$ using Eqs.~(\ref{Eq:W1 and W2}), (\ref{eq:solve f}), and (\ref{Eq:pfct});
   \item Determine $\sigma(t)$ using Eqs.~(\ref{Eq:sigmainv}) and (\ref{Eq:sigma});
   \item Compute the output vector $\mathbf{y}$ as well as the history vectors $\mathbf{i}^\textit{hist}_{\textrm{ac}}$ and $\mathbf{i}^\textit{hist}_{\textrm{dc}}$ for the next time-point using Eq.~(\ref{Eq:W}).
\end{enumerate}

This algorithm is referred to as the Two-Stage Algorithm because it involves two Matrix-Vector Multiplications (MVMs): Step 1) is performed by combining Eqs.~(\ref{Eq:W1 and W2}), (\ref{eq:solve f}) into a single MVM following which Eq.~(\ref{Eq:pfct}) is used to determine $\sigma_\textit{rec}$. Step 2) is a simple binary word concatenation and is therefore instantaneous. Step 3) constitutes the second stage MVM of the algorithm. 

\subsection{Low-latency Implementation of the DMM}

\begin{figure}[!t]
    \centering
    \includegraphics[width=3.4in]{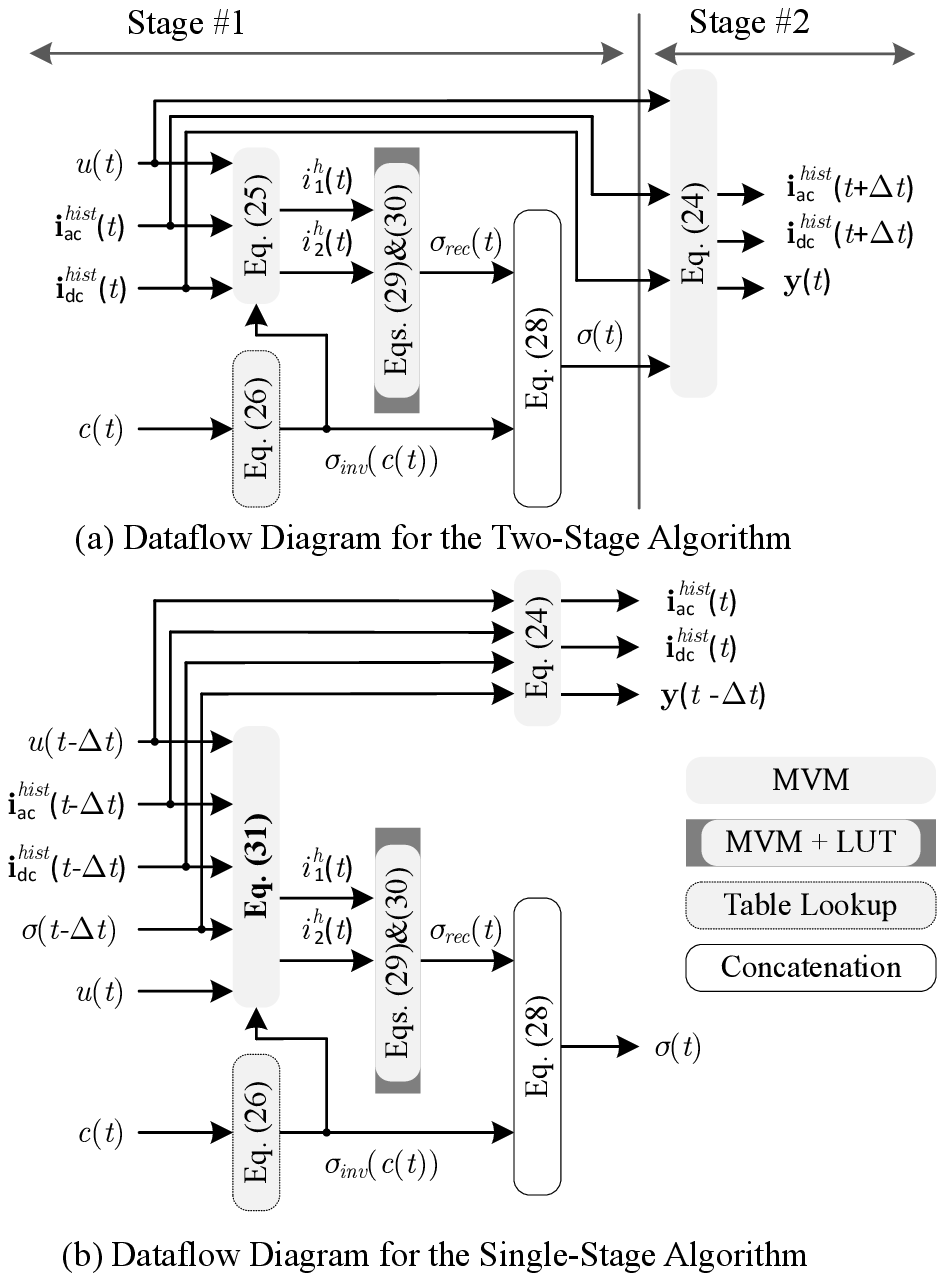}
    \caption{Data-flow Diagram: (a) Two-Stage Algorithm; (b) Single-Stage Algorithm.}
    \label{fig:dataflow}
\end{figure}

Fig.~\ref{fig:dataflow}.a illustrates a dataflow diagram for the Two-Stage Algorithm, and shows the data dependency of the two stages (through $\sigma(t)$) that forces their serial execution. Moreover, each stage involves matrix vector multiplications (MVMs) that hinders reaching to a small time-step. 

It is possible to  reduce the time-step if one rewrites Eqs.~(\ref{Eq:W1 and W2}) in the following form:

\begin{equation}   
\left\{
\begin{array}{llll}
 i^h_1(t) & = & 
 \mathbf{H}^{\sigma_\textit{inv}(c(t))}_{1\textrm{,u}}
 &u(t) \\
                   & + & 
 \mathbf{H}^{\sigma_\textit{inv}(c(t))}_{1\textrm{,ac}}
 \mathbf{H}^{\sigma(t - \Delta t)}_\textrm{ac,u}
 &u(t - \Delta t) \\
                   & + & 
 \mathbf{H}^{\sigma_\textit{inv}(c(t))}_{1\textrm{,ac}}
 \mathbf{H}^{\sigma(t - \Delta t)}_\textrm{ac,ac}
 &\mathbf{i}^\textit{hist}_\textrm{ac}(t - \Delta t) \\
                   & + & 
 \mathbf{H}^{\sigma_\textit{inv}(c(t))}_{1\textrm{,ac}}
 \mathbf{H}^{\sigma(t - \Delta t)}_\textrm{ac,dc}
 &\mathbf{i}^\textit{hist}_\textrm{dc}(t - \Delta t) \\
 i^h_2(t) & = & 
 \mathbf{H}_2
 \mathbf{H}^{\sigma(t - \Delta t)}_\textrm{dc,u}
 &u(t - \Delta t) \\
                   & + & 
 \mathbf{H}_2
 \mathbf{H}^{\sigma(t - \Delta t)}_\textrm{dc,ac}
 &\mathbf{i}^\textit{hist}_\textrm{ac}(t - \Delta t) \\
                   & + & 
 \mathbf{H}_2
 \mathbf{H}^{\sigma(t - \Delta t)}_\textrm{dc,dc}
 &\mathbf{i}^\textit{hist}_\textrm{dc}(t - \Delta t)
\end{array}
\right..
  \label{Eq:W1 and W2 LL}
\end{equation}

Hence, knowing $u(t-\Delta t)$, $\sigma(t-\Delta t)$, $\mathbf{i}^\textit{hist}_\textrm{ac}(t-\Delta t)$, $\mathbf{i}^\textit{hist}_\textrm{dc}(t-\Delta t)$, $u(t)$, and $c(t)$ would suffice to determine $\sigma(t)$. Fig.~\ref{fig:dataflow}.b illustrates how this rewriting yields a Single-Stage Algorithm version of the DMM by breaking the data dependency. The only limitation of this approach is that two consecutive simulation steps are needed to produce the output vector $\mathbf{y}(t)$. Section \ref{sec:FPGA} will demonstrate that the hardware implementation of the Single-Stage results in an input-output latency of two time-steps, but that it is legitimate since the simulation time-step is halved compared to the Two-Stage version of the algorithm.

\section{FPGA Implementation}
\label{sec:FPGA}

\begin{figure}[!t]
\centering
\includegraphics[width=3.3 in]{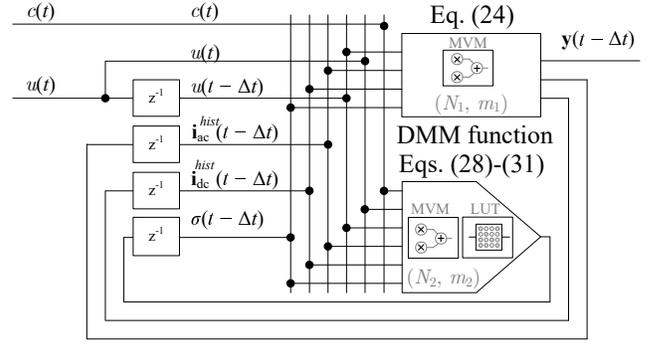}
\caption{Datapath of the hardware implementation of the LLC simulator using Single-Stage DMM Algorithm.}
\label{fig:Hardware architecture}
\end{figure}

\begin{figure}[!t]
\centering
\includegraphics[width=3.3 in]{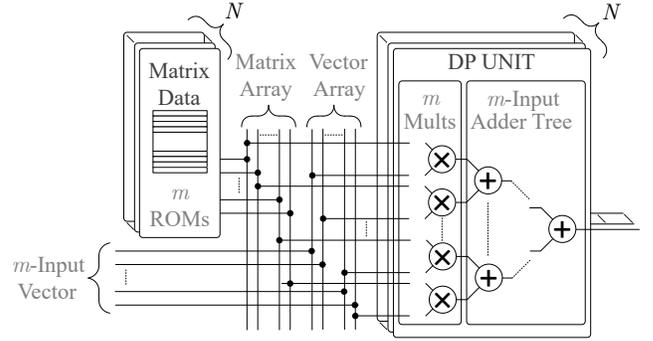}
\caption{The $(N, m)$ MVM module used in the proposed LLC simulator: The module is composed of $N\times m$ ROMs and $N$ $m$-Input DP units. }
\label{fig:MVM}
\end{figure} 

\subsection{Hardware Implementation}

A  hardware implementation of the Single-Stage DMM algorithm datapath is shown in Fig.~\ref{fig:Hardware architecture}. The architecture is almost a one to one map of the dataflow diagram of Fig.~\ref{fig:dataflow}.b, and  consists of two main computing units: a)~The first unit is devoted to updating history terms and computing outputs of interest, i.e. Eq.~(\ref{Eq:W}); b)~The second unit evaluates the DMM function, i.e.  Eqs.~(\ref{Eq:sigma})-(\ref{Eq:W1 and W2 LL}). Eq.~(\ref{Eq:W}) is implemented by a dedicated MVM module; Eqs.~(\ref{eq:solve f}) and (\ref{Eq:W1 and W2 LL}) are combined and implemented by a dedicated MVM module as well. Eqs.~(\ref{Eq:pfct}) is a simple lookup table. Eq.~(\ref{Eq:sigma}) consists of a bit string concatenation and comes at no hardware cost. 

Each MVM module from Fig.~\ref{fig:Hardware architecture} is shown in Fig.~\ref{fig:MVM}. The MVM module implements the multiplication of an $n\times m$ matrix by an $m\times 1$ vector, and is made up of a set of $N\times m$ Read-Only Memories (ROMs) and $N$ $m$-input Dot-Product (DP) units, where $1<N<n$ is a parallelization parameter.

The two MVM modules present in the LLC simulator of Fig.~\ref{fig:Hardware architecture} handle $n_1\times m_1$ = $7\times 5$ and $n_2\times m_2$ = $4\times 6$ matrices, and as such are defined by parameters $(N_1, m_1)$ and $(N_2, m_2)$. In Section~\ref{sec:design space}, a design space exploration is presented to discuss the impact of parameters $N_1$ and $N_2$ on area occupation, and simulation time-step.

\subsection{Number Format}

An FPGA can handle real arithmetic using either fixed-point (FXP) or floating-point (FP) format. The FXP format uses less hardware and yields a datapath of lower latency, but has a limited dynamic range. The FP number format allows for a larger dynamic range, but its hardware arithmetic operators are costly in terms of FPGA resource consumption, and require deeper pipelines than their FXP counterparts. Due to latency considerations, this paper considers solely the FXP number format. The limited dynamic range issue of the FXP format is addressed by normalizing matrix entries using a per-unit scale.

The targeted FPGA, a Kintex K325T from Xilinx, uses an asymmetric multiplication block ($25\times 18$ signed). Hence, the selected number format used by the LLC simulator will be different whether we are dealing with a vector (FXP~25.23) or a matrix (FXP~35.29) in order to offer a good precision to the precomputed matrices and good computation accuracy, while reducing the area footprint of the simulator. A similar idea was presented in \cite{Blanchette2012} and is adopted here.

\begin{table}[!t]
\begin{center}
\caption{design exploration targetting the Kintex K325T} 
\begin{tabular}{|l|l|r|r|r|}

\hline \hline

& Item & 40 MHz & 200~MHz & 320~MHz \\
\hline  \hline

\parbox[t]{2mm}{\multirow{5}{*}{\rotatebox[origin=c]{90}{FPI}}} 
& Min. Time-Step & 25~ns     & 25~ns      & 25~ns \\
& In-Out Latency & 50~ns     & 50~ns      & 50~ns \\
& Registers      & 1,224 (0.3\%) & 1,697  (0.4\%) & 2,373 (0.6\%) \\
& LUTs           & 1,090 (0.5\%) & 1,095  (0.5\%) & 1,223 (0.6\%) \\
& DSP Blocks     & 88 (10.5\%)   & 88 (10.5\%)    & 88 (10.5\%) \\
& BRAM           & 22 (4.9\%)    & 22  (4.9\%)    & 22 (4.9\%) \\

\hline \hline

\parbox[t]{2mm}{\multirow{5}{*}{\rotatebox[origin=c]{90}{HRT}}} 
& Min. Time-Step & 100~ns  & 40~ns   & 34.375~ns \\
& In-Out Latency & 200~ns  & 80~ns   & 68.750~ns \\
& Registers      & 211 (0.1\%) & 350 (0.1\%) & 498 (0.1\%) \\
& LUTs           & 277 (0.1\%) & 329 (0.2\%) & 311 (0.2\%) \\
& DSP Blocks     & 22 (2.6\%)  & 22 (2.6\%)  & 22 (2.6\%) \\
& BRAM           & 5.5 (1.2\%) & 5.5 (1.2\%) & 5.5 (1.2\%) \\

\hline \hline

\end{tabular}

\label{Table:FPGA Usage} 
\end{center}
\end{table}

\subsection{Design Space Exploration}
\label{sec:design space}

This section presents a design space exploration whose purpose is to evaluate the impact of parameters $N_1$ and $N_2$ on area occupation, and simulation time-step. Two implementations are considered in this design exploration, the first approach consists in a Fully Parallel Implementation (FPI) that sets $N_1=n_1=7$, and $N_2=n_2=4$; whereas the second approach will resort to a Hardware Reuse Technique (HRT) that consists in setting $N_1<n_1$ and $N_2<n_2$, thus time-multiplexing the computations. In this paper, the adopter HRT approach sets $N_1=1$ and $N_2=1$.

Also considered is the effect of the FPGA clock frequency on the timing performance and simulation time-step. Various clock frequencies are investigated, namely 40~MHz, 200~MHz and 320~MHz. The timing closure is met by varying the pipelining depth of the DP Unit accordingly. When the 40~MHz clock frequency is targeted, the DP Unit is purely combinational.

Table~\ref{Table:FPGA Usage} presents the design exploration results for the Kintex K325T for each design approach and each target frequency. It shows that the smallest simulation time-steps (25~ns) are obtained when a FPI approach is adopted. FPI is however the most expensive approach in terms of hardware consumption. The HRT trades simulation time-step for area footprint. Hence, this approach allows considerable saving in hardware utilization (up to 4 folds), with a very acceptable impact on the simulation time-steps, which are less or equal to 100~ns. For HRT, smaller time-steps are achieved by increasing the clock frequency, with the smallest time-step (34.375~ns) obtained for the highest clock frequency (320~MHz). The 200~MHz implementations are the most balanced options for both FPI and HRT approaches. 

That being said, all reported implementations are characterized by a very low hardware utilization, a small simulation time-step ($\le$ 100~ns), and an input-output (In-Out) latency of two time-steps. These notable results are more obvious from Table~\ref{tab:FPGAcmp1}, which lists from the literature works dealing with the real-time simulation of high-frequency converters, i.e. $\ge$~50~kHz. As one can see from Table~\ref{tab:FPGAcmp1}, where the 200~MHz FPI and HRT results have been reproduced, the proposed implementations have indeed a very small footprint and offer one of the smallest time-steps ever reported in the literature. It is noteworthy that, thanks to the proposed DMM, these outcomes are obtained using an implicit solver, without decoupling any part of the circuit, and while using a simultaneous and exact switch state solution.

\begin{figure}[!t]
\centering
\includegraphics[width=3.3 in]{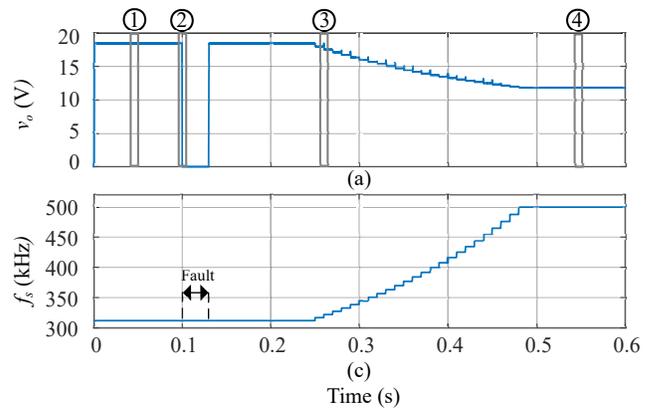}
\caption{LLC with Parameter Set \#2: (a): Output voltage, $v_o$, (b): Switching frequency ($f_s$) and fault period.}
\label{fig:outputresult-case1}
\end{figure}

\begin{figure*}[!t]
\centering
\includegraphics[width=7 in]{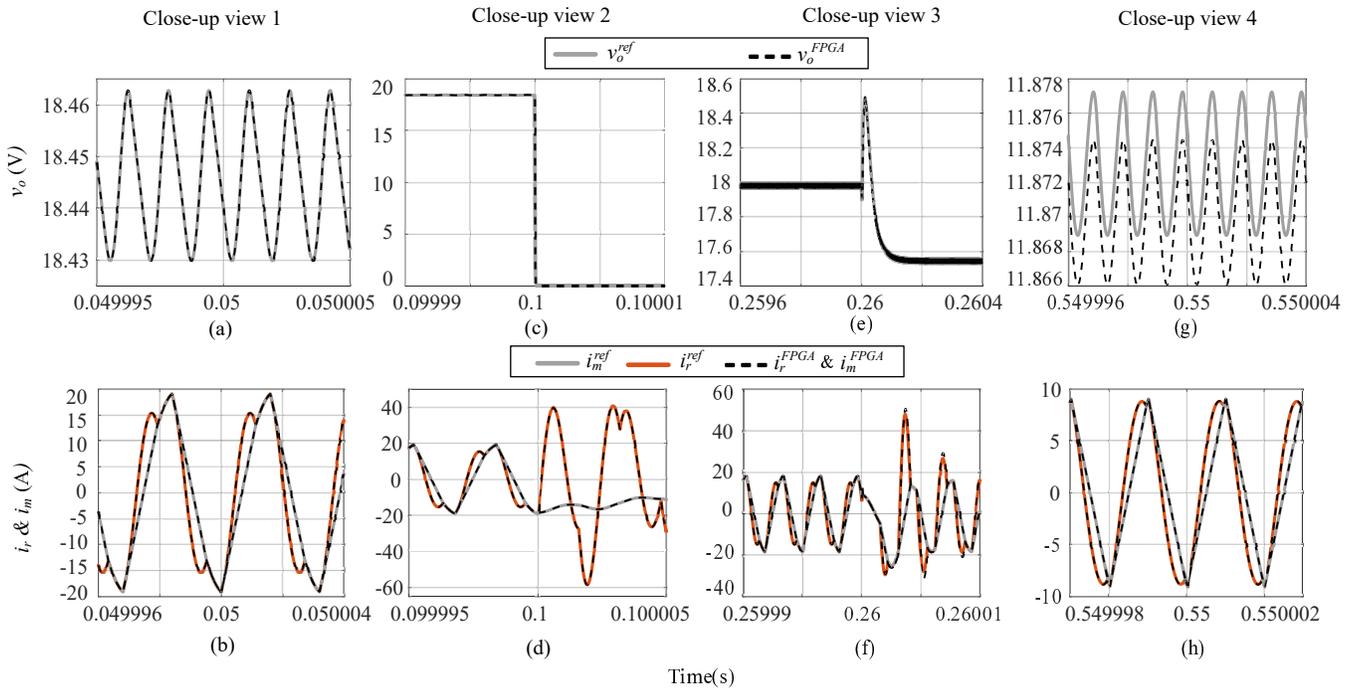}
\caption{LLC with Parameter Set \#2: Close-up views of $v_o$ and $i_r, i_m$. (a)-(b): Close-up view~1; (c)-(d): close-up view~2; (e)-(f): close-up view~3; and (g)-(h): close-up view~4. Output voltage close-up views are shown in a wider time-span than resonant and magnetizing currents.}
\label{fig:zoomedoutputresult-case1}
\end{figure*}

\begin{table*}[!t]
\centering
\caption{Comparison with the existing work on the FPGA-based real-time simulation of resonant converter.}
\label{tab:FPGAcmp1}
\begin{adjustbox}{width=7.1in}
\begin{tabular}{|c|c|c|c|c|c|c|c|c|c|c|c|}

\hline
\multicolumn{5}{|c|}{Power Electronic Circuit}  & \multicolumn{7}{c|}{FPGA Resource Consumption} \\ \hline

~& 
$\Delta t$ (ns)  & 
Solver & 
Application & 
Sw. Frq. (kHz) & 
FPGA & 
Clk (MHz) & 
Number Format & 
LUTs & 
Registers & 
BRAM & 
DSP48 \\ \hline

FPI & 
25 & 
BE & 
LLC & 
500 & 
K7-325\textsuperscript{1} & 
320 & 
FXP 32.29/25.22 & 
1,095 (0.5\%) & 
1,697 (0.4\%) & 
22 (4.9\%) & 
88 (1.05\%) \\ \hline

HRT & 
40 & 
BE & 
LLC & 
500 & 
K7-325\textsuperscript{1} & 
320 & 
FXP 32.29/25.22 & 
329 (0.2\%) & 
350 (0.1\%) & 
5.5 (1.2\%) & 
22 (2.6\%) \\ \hline

\cite{FeiGaoLLC2019} & 
15 & 
FE & 
Batt. Charger & 
160 & 
K7-410\textsuperscript{2} & 
66.67 & 
FXP 25.23 & 
45,560 (17.9\%) & 
44,277 (8.7\%) & 
106 (13.3\%) & 
43 (2.8\%) \\ \hline 

\cite{Hadizadeh2019} & 
36 & 
BE & 
AC-DC-AC & 
50 & 
V7-485\textsuperscript{3} & 
200 & 
SFP\textsuperscript{4} & 
116,670 (38\%) & 
68,920 (11\%) & 
1,202 kb (3\%) & 
 762 (27\%)  \\ \hline

\cite{JiLLC2016} & 
100 & 
FE & 
LLC & 
60 & 
NR\textsuperscript{6} & 
NR & 
NR & 
NR & 
NR & 
NR & 
NR \\ \hline

\cite{LIU2019} & 
40 & 
PC\textsuperscript{5}: FE-BE & 
NPC & 
NR & 
K7-410\textsuperscript{2} & 
25 & 
32 & 
47,028 (25.8\%) & 
42,642 (11.2\%) & 
91 (11.6\%) & 
120 (17.7\%) \\ \hline

\cite{RealTimeBattryCharger2} & 
40 & 
Sw. fct + FE & 
Batt. charger & 
100  & 
ZYNQ & 
200 & 
FXP 32.20 & 
NR & 
NR & 
NR & 
NR  \\ \hline  

\cite{Blanchette2012} & 
80 & 
BE & 
3-$\Phi$ inverter & 
200 & 
V5-50\textsuperscript{7} & 
200 & 
FXP 35.30/25.16 & 
229 (7\%) & 
3,531 (10.8\%) & 
44 (33.3\%) & 
176 (61.1\%) \\ \hline

\cite{Benigni2017} & 
50 & 
Sw. fct + FE & 
3-$\Phi$ inverter & 
100 & 
V7-485\textsuperscript{3} &  
200 & 
FXP 72.43 & 
3,943 (1.3\%) & 
893 (0.1\%) &
NR &
170 (6.1\%) \\ \hline

\multicolumn{12}{|l|}{}\\ [-0.5em]

\multicolumn{12}{|l|}{
\textsuperscript{1}K7-325: Kintex XC7K325T. 
\textsuperscript{2}K7-410: Kintex XC7K410T.
\textsuperscript{3}V7-485: Virtex XC7VX485T
\textsuperscript{4}SFP: Single precision FP.
\textsuperscript{5}PC: Predictor-Corrector.
\textsuperscript{6}NR: Not Reported. .
\textsuperscript{7}V5-50: Virtex VC5VSX50T.
} \\  [0.5em] \hline

\end{tabular}
\end{adjustbox}
\end{table*}

\subsection{Computational Accuracy}

The computational accuracy of the hardware LLC simulator (200 MHz FPI) is assessed through a test sequence lasting 0.6s. Parameter Set \#2 in Table~\ref{tab:LLCDATA} is considered. During the test sequence, the input voltage is kept constant at 400~V. The FPGA simulation results are validated against an offline iterative solution, as shown in Fig.~\ref{fig:outputresult-case1}, where the output voltage ($v_o$) and the switching frequency are shown. The test sequence comprises the following steps:

\begin{enumerate}
    \item $t=0$s-$0.1$s: Operation of the inverter at $f_s=312.5$~kHz.
    \item $t=0.1$s-$0.13$s: Load is shorted (fault); $f_s=312.5$~kHz.
    \item $t=0.13$s-$0.25$s: Fault cleared; $f_s=312.5$~kHz.
    \item $t=0.25$s-$0.48$s: The switching frequency $f_s$ is increased from $312.5$~kHz to $500$~kHz.
    \item $t=0.48$s-$0.6$s: Operation at $f_s=500$~kHz.
\end{enumerate}

From Fig.~\ref{fig:outputresult-case1}, one can see that the $v_o$ gradually decreases and finally settles at $\approx$ 12~V as $f_s$ is increased from $312.5$~kHz to $f_s = 500$~kHz. Fig.~\ref{fig:zoomedoutputresult-case1} offers close-up views of four instants, as identified in Fig.~\ref{fig:outputresult-case1}. The FPGA results are overlapped in the same figure and show very good agreements with the reference. At 500~kHz, the output voltage shows a slight dc offset (Fig.~\ref{fig:zoomedoutputresult-case1}.g) that is not larger that 0.2\%.

\begin{table}[!t]
\centering
\caption{LLC with Parameter Set \#2\cite{Fei2018}: 2-Norm Relative Errors}
\label{tab:2norm}
\begin{tabular}{|c|c|c|c|}
\hline
Sequence    & $v_o$ & $i_r$ & $i_m$ \\ \hline \hline
0.00s-0.10s &  0.024\% & 0.383\% & 0.120\% \\
0.10s-0.13s &  0.001\% & 0.001\% & 0.016\% \\
0.13s-0.25s &  0.021\% & 0.338\% & 0.131\% \\
0.25s-0.48s &  0.006\% & 0.391\% & 0.131\% \\
0.48s-0.60s &  0.023\% & 0.499\% & 0.048\% \\ \hline
0.00s-0.60s &  0.018\% & 0.336\% & 0.124\% \\
\hline
\end{tabular}
\end{table}

To further assess the accuracy of the FPGA result, each signal is compared to its offline reference. The 2-norm relative error is used as a measure of accuracy~\cite{Gautschi2Norm}:

\begin{equation}
e = \dfrac{ ||\mathbf{f}_s-\mathbf{f}_r||_2}{||\mathbf{f}_{r}||_2}
\label{norm2}
\end{equation}
where $\mathbf{f}_s$ and $\mathbf{f}_r$ are FPGA results and offline reference, respectively. Table \ref{tab:2norm} reports the 2-norm errors for the FPGA results for each sub-sequence as well as for the entire test sequence. All the reported errors are below 0.5\% and show the very good performance of the FPGA implementation in different operating mode and condition of the LLC.

\section{Conclusion}

This paper presented a direct mapped method for the accurate real-time simulation of high switching frequency resonant converters. The method obviates the need for iterations while providing accurate simulation results. The benefits of the proposed method was demonstrated through the implementation of an FPGA-based LLC real-time simulator capable of achieving a time-step of 25~ns while offering a very small hardware footprint. A design space exploration was proposed to discuss the opportunity of further reducing the area occupation by reusing parts of the computational datapath. It was shown that such an approach can result in considerable savings for higher clock frequencies. The LLC simulator performance has been evaluated for the case of a 500 kHz LLC converter. It has been shown that the DMM real-time results are in excellent agreement with those obtained by offline iterative solution. A 2-norm relative error of less than 0.5\% for various operating modes has been reported, which was also shown to hold true in the presence of a fault.

\bibliographystyle{IEEEtranTIE}
\bibliography{IEEEabrv,referencelib} 

\begin{IEEEbiographynophoto}
{Hossein Chalangar}received the B.Sc. and M.Sc degrees in electrical engineering from K. N. Toosi University of Technology (KNTU), Tehran, Iran, in 2011 and
2013, respectively. He is currently working toward the Ph.D. degree at Polytechnique Montr\'eal, Canada. His research interests include real-time simulation of power systems and power electronics.
\end{IEEEbiographynophoto}

\begin{IEEEbiographynophoto}
{Tarek Ould-Bachir}
(M'08) received the M.A.Sc. and Ph.D. degrees in electrical engineering from the Polytechnique Montr\'eal, Montreal, QC, Canada, in 2008 and 2013, respectively. From 2007 to 2018, he was with OPAL-RT Technologies, holding various positions in the R\&D department. From 2018 to 2020, he was a Research Associate with Polytechnique Montr\'eal. He is currently an Assistant Professor with Polytechnique Montr\'eal. 
\end{IEEEbiographynophoto}

\begin{IEEEbiographynophoto}
{Keyhan Shesheykani} (M’10-SM’13) received the
B.S. degree in electrical engineering from Tehran
University, Tehran, Iran, in 2001, and the M.S.
and Ph.D. degrees in electrical engineering from
Amirkabir University of Technology (Tehran Polytechnique), Tehran, Iran in 2003 and 2008, respectively. He was with Ecole Polytechnique F\'ed\'erale
de Lausanne, Lausanne, Switzerland, in September
2007 as a Visiting Scientist and later as a Research
Assistant. From 2010 to 2016, he was with Shahid
Beheshti University, Tehran. He was an Invited Professor at the EPFL from June to September 2014. He joined the Department
of Electrical Engineering, Polytechnique Montr\'eal, Montreal, QC, Canada, in
2016, where he is currently an Associate Professor. Dr. Sheshyekani currently
serves as the Associate Editor of IEEE Transactions on Electromagnetic
Compatibility. His research interests include power system modeling and
simulation, smart grids and electromagnetic compatibility.
\end{IEEEbiographynophoto}

\begin{IEEEbiographynophoto}
{Jean Mahseredjian}
(F'13) received the M.A.Sc. and Ph.D. degrees in electrical engineering from the \'Ecole Polytechnique de Montr\'eal, Montreal, QC, Canada, in 1985 and 1991, respectively. From 1987 to 2004 he was with IREQ (Hydro-Québec), Québec, Canada, working on research-and-development activities related to the simulation and analysis of electromagnetic transients. In 2004, he joined the faculty of electrical engineering at \'Ecole Polytechnique de Montréal.
\end{IEEEbiographynophoto}

\vfill

\end{document}